\documentclass{emulateapj}

\usepackage{graphicx,natbib}
\usepackage{color}
\usepackage{float}
\usepackage{txfonts}

\begin{document}

   \title{\textbf{V348~And, and V572~Per: bright triple systems with eccentric eclipsing binaries}\thanks{Based on the data from 2-m
   telescope at the Ond\v{r}ejov observatory in Czech Republic.}}

 \shorttitle{V348~And, and V572~Per: bright triple systems with eccentric eclipsing binaries}
 \shortauthors{Zasche et al.}

   \author{P. Zasche\altaffilmark{1}, 
          R. Uhla\v{r}\altaffilmark{2},
          P. Svoboda\altaffilmark{3},
          J. Jury\v{s}ek\altaffilmark{4},
          D. Kor\v{c}\'akov\'a\altaffilmark{1},
          M. Wolf\altaffilmark{1},
          M. \v{S}lechta\altaffilmark{5},
          L. Kotkov\'a\altaffilmark{5}}

\email{zasche@sirrah.troja.mff.cuni.cz}

 \affil{
  \altaffilmark{1} Astronomical Institute, Charles University, Faculty of Mathematics and Physics, CZ-180~00, Praha 8,
             V~Hole\v{s}ovi\v{c}k\'ach 2, Czech Republic,\\ 
  \altaffilmark{2} Private Observatory, Poho\v{r}\'{\i} 71, CZ-254 01 J\'{\i}lov\'e u Prahy, Czech Republic \\
  \altaffilmark{3} Private Observatory, V\'ypustky 5, CZ-614 00, Brno, Czech Republic\\
  \altaffilmark{4} Institute of Physics, The Czech Academy of Sciences, Na Slovance 1999/2, CZ-182 21, Praha, Czech Republic\\
  \altaffilmark{5} Astronomical Institute, The Czech Academy of Sciences, CZ-251 65, Ond\v{r}ejov, Czech Republic\\
  }

  \date{Received 2019 April 2; revised 2019 June 3; accepted 2019 June 25; published 2019 August 2}

\begin{abstract}
 \noindent The eclipsing binaries are still important objects for our understanding of the Universe.
Especially these ones located within the more complex multiple systems can help us solving the
problem of their origin and subsequent evolution of these higher-order multiples. Photometry and
spectroscopy spanning over more than 25 years were used for the first complete analysis of the two
bright triple systems, namely V348~And and V572~Per.
  The light curves in photometric filters were combined together with the radial velocities and
analysed simultaneously, yielding the precise physical parameters of the eclipsing components of
these multiple systems.
  The system V348 And consists of two eclipsing components with its orbital period of about 27.7
days. The system is a very detached one, and both eclipses are rather narrow, lasting only about
0.016 of its period. The visual orbit of the wide pair has the period of about 87 years. All three
components of the system are probably of B8-9 spectral type, and the parallax of the system was
slightly shifted to the value of 2.92~mas. On the other hand, the system V572 Per shows apsidal
motion of its inner orbit, the orbital period being of about 1.2 days, while the apsidal motion of
about 48 years. The components are of A and F spectral types, while the motion with the third
component around a common barycenter is only negligible. According to our modelling, this system is
not a member of open cluster Alpha~Persei.
 \end{abstract}



 \section{Introduction}

The eclipsing binaries still represent the most general method how to derive the stellar masses,
radii and luminosities most precisely. The fruitful combination of the observational techniques
like photometry and spectroscopy is still being used also for deriving the temperatures, surface
gravities or limb darkening, but also to compute the distances to these systems.

On the other hand, studying the binaries as parts of the higher order multiple systems can bring us
new important results connected with the stellar origin and evolution. We can ask -- how many
multiple systems are there within the stellar population? What is the multiplicity fraction of the
field stars, and is this number still the same? Or is it somehow evolving in time and the
multiplicity fraction can be tracked as different between PopI and PopIII stars? Is the mass ratio,
period ratio, or eccentricity ratio the same for the low mass as well as for the high mass stars?
What is the role of the Kozai cycles (see e.g. \citealt{2001ApJ...562.1012E})? Can the effects like
synchronization or circularization predicted by the tidal theories \citep{2008EAS....29...67Z} be
traced in particular systems via deriving their orbital and physical parameters? These and many
other still open questions play a crucial role in our theories of stellar formation and evolution,
see e.g. \cite{2008MNRAS.389..925T}, \cite{2003A&A...397..159H}, or \cite{2014AJ....147...87T}. And
of course, the models can be tested and verified only when using the real data, which can be
obtained only via studying the particular system and derive its parameters. The importance of
dedicated studies of particular systems with higher hierarchies was presented e.g. in the updated
version of the Multiple Star Catalog MSC, \cite{2018ApJS..235....6T}. This is the main aim of the
present paper, to bring new results on two new multiple systems never studied before. And moreover,
both these systems are bright enough for subsequent follow-up observations.

Due to this reason, we have focused our effort on two systems for which their light curves
(hereafter LC) and radial velocity curves were not studied yet (namely V348~And, and V572~Per).
Besides the inner eclipsing pair, both these systems also contain the distant third component
detected via interferometry with its rather long orbital period. Moreover, both these stars show
the eccentric orbits.

\section{The analysis}

Whole our analysis is using a classical combination of the photometry and spectroscopy into one
joint approach. If we combine these methods, we can obtain the physical parameters of both
eclipsing components as well as the parameters of their mutual orbit. As a consequence, having the
complete information about their masses, inclinations, periods, etc. we can also fill in still
quite incomplete statistics of the multiple (triple and quadruple) systems. All of these
distributions of orbital and physical parameters are being used for discussions about the origin
and subsequent evolution of such multiples (\citealt{2008MNRAS.389..925T} and
\citealt{2014AJ....147...87T}).

All of our new spectroscopic observations were secured in the Ond\v{r}ejov observatory in Czech
Republic, using the 2-meter telescope. The classical slit spectrograph has its resolution R $\sim$
12500. The individual exposure times were chosen according to the quality of the particular night,
typically 800--3600 seconds. All the spectrograms were reduced in a standard way, the wavelength
calibration was made via a ThAr comparison spectra obtained before and after the stellar ones. The
flatfields were taken in the beginning and end of the night and their averages were then used for
the reduction.
 After then, the radial velocities (hereafter RV) were derived with the program SPEFO
(\citealt{1996A&A...309..521H}, or \citealt{1996ASPC..101..187S}), on several absorption lines in
the measured spectral region around H$_\alpha$ (usually \emph{Fe}, \emph{Ca}, or \emph{Si} lines),
with using the zero point correction via measuring the telluric lines.

Owing to relatively high brightness of these stars, only rather small telescopes were used for the
photometric observations. The system V348~And was observed (by PS) with only the 34-mm refractor at
his private observatory in Brno, Czech Republic, using the SBIG ST-7 CCD camera. The second star
V572~Per was monitored with the similar instrument at the private observatory (by RU) in
J\'{\i}lov\'e u Prahy, Czech Republic, using a G2-0402 CCD camera. All the measurements were
reduced in a standard way using the programme C-MUNIPACK\footnote{See
http://c-munipack.sourceforge.net/} which is based on aperture photometry and uses the standard
DAOPHOT procedures \citep{1993ASPC...52..173T}. The photometric data were obtained during the time
span 2007--2018. Nevertheless, some of the older data were only used for the minima times
derivation. All of these data were secured in the Johnson-Cousins photometric system
\citep{1990PASP..102.1181B}, particularly the system V348~And was observed in $BVR$, while the
system V572~Per in $BVRI$ filters.

 \begin{figure*}
   \centering
   \includegraphics[width=0.99\textwidth]{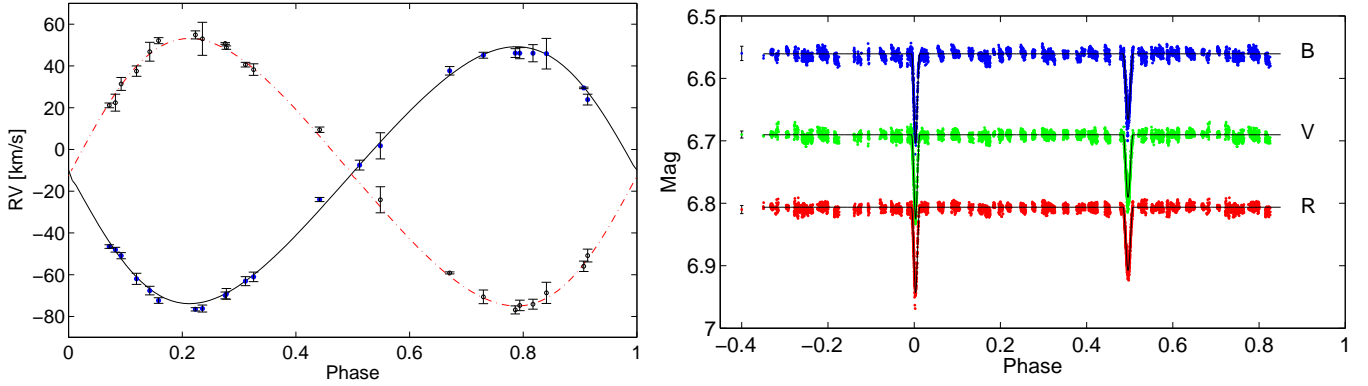}
   \caption{Results of our LC+RV analysis of V348 And as based on the {\sc PHOEBE} fitting. \emph{Left:} For the
   radial velocities, the primary is plotted as a solid line (and full dots), while secondary as a dashed line
   (and open circles). \emph{Right:} For the photometry the data collected during more than 10 years. Individual
   curves were shifted in y-axis for better brevity of the picture. Typical precision of individual observations
   are also plotted as short error bars on the left.}
   \label{FigRVLC_V348And}
  \end{figure*}

  \begin{figure}
   \centering
   \includegraphics[width=0.49\textwidth]{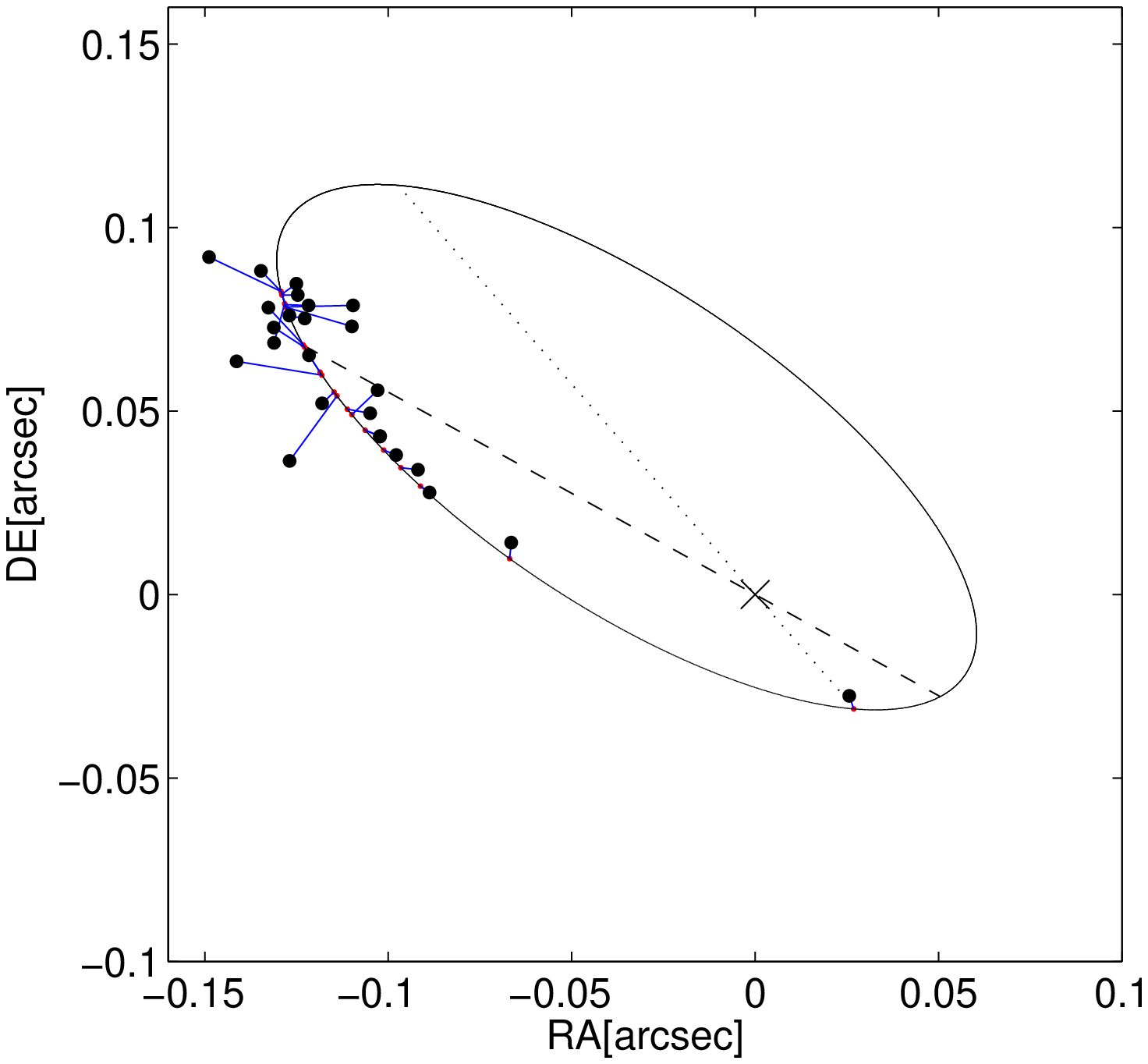}
   \caption{Visual orbit of V348 And. The individual observations are connected with their
   theoretical positions on the orbit, while the dotted line stands for the line of the apsides and
   the dashed line stands for the line of the nodes.}
   \label{FigOrbit_V348And}
  \end{figure}

Both photometric and spectroscopic observations were studied in the standard approach. Hence, the
program {\sc PHOEBE} \citep{2005ApJ...628..426P}, which is using the classical Wilson-Devinney
algorithm (\cite{1971ApJ...166..605W} and its later modifications) was used for the analysis. It
allows us to fit the relevant physical parameters of the eclipsing components, as well as their
relative orbit. For the analysis, we used several assumptions. At first, the primary temperature
was set to the value corresponding to the particular spectral type (see calibrations by
\cite{2013ApJS..208....9P} and the more updated web
site\footnote{www.pas.rochester.edu/$\sim$emamajek/EEM$\_$dwarf$\_$UBVIJHK$\_$colors$\_$Teff.txt}).
The coefficients of the limb-darkening were interpolated from van Hamme's tables
\citep{1993AJ....106.2096V}. The coefficients of albedo $A_i$, and the gravity darkening
coefficients $g_i$ were fixed at their suggested values. And finally, the third light was also
computed because we deal with the triple systems and the additional light can be easily attributed
to this distant component.

The errors of individual parameters were derived via combined approach using besides the PHOEBE
also the code called {\sc Pyterpol}\footnote{https://github.com/chrysante87/pyterpol/wiki}
\citep{2016A&A...594A..55N}. It derives the radiative and kinematic parameters of components via
comparison of observed spectra to the synthetic ones.

For the fitting of apsidal motion and visual orbit (see below) our own codes were used. These
programs are using the weighted least squares method and the simplex algorithm. For error
estimation the boot-strapping method was used.

\section{V348~And}

The first system is the northern-hemisphere star V348~And (= HIP~1233, HD~1082, $V_{max}$ =
6.6~mag). Since \cite{1935sgcs.book.....D} its spectral type is usually being classified as B9.
Despite its brightness and northern declination, its detailed analysis is still missing. The star
was classified as an eclipsing binary based on the Hipparcos data \citep{HIP}, having the supposed
orbital period of about 5.5~days. However, according to our observations obtained during 2007--2008
we found there no evidence of eclipse. Hence, we prematurely stated that the star was incorrectly
classified as an eclipsing binary, see our previous paper on this star in
\cite{2008IBVS.5827....1Z}.

According to our new findings presented in this paper, we found that the star is eclipsing, but
with much longer orbital period. Moreover, the star is also known as a visual binary, having the
distant component of about 0.1$^{\prime\prime}$ away from the primary. However, its orbital period
is still rather uncertain (see below). The parallax of the system was derived as 3.02~mas
\citep{2007A&A...474..653V}, while more recently by GAIA \citep{2018A&A...616A...1G} as 2.16~mas.

\begin{table*}
  \caption{The parameters from the LC+RV fitting of both systems.}
  \label{LCRVparam}
  \centering
  \begin{tabular}{c | c c c | c c c}
\hline \hline
                   & \multicolumn{3}{c|}{V348 And} & \multicolumn{3}{c}{V572 Per} \\  
Parameter          & Primary & Secondary & Tertiary                          &  Primary & Secondary & Tertiary \\
 \hline
 $HJD_0$           & \multicolumn{2}{c}{2455923.129 $\pm$ 0.025 } & --       &  \multicolumn{2}{c}{2457007.3701 $\pm$ 0.0017 } & -- \\
 $P$ [d]           & \multicolumn{2}{c}{27.703514 $\pm$ 0.000325} & --       &  \multicolumn{2}{c}{1.2131789 $\pm$ 0.0000006} & -- \\
 $a$ [R$_\odot$]   & \multicolumn{2}{c}{68.28 $\pm$ 0.87 }  & --             &  \multicolumn{2}{c}{7.01 $\pm$ 0.07}       & -- \\
 $v_\gamma$ [km/s] & \multicolumn{2}{c}{-11.63 $\pm$ 0.52 } & --             &  \multicolumn{2}{c}{-0.05 $\pm$ 0.20 }    & -- \\
 $e$               & \multicolumn{2}{c}{0.116 $\pm$ 0.012}  & --             &  \multicolumn{2}{c}{0.029 $\pm$ 0.002}     & -- \\
 $\omega$ [deg]    & \multicolumn{2}{c}{95.6 $\pm$ 0.2}     & --             &  \multicolumn{2}{c}{10.0 $\pm$ 0.1}        & -- \\
$\dot\omega$ [deg/yr] &\multicolumn{2}{c}{--}               & --             &  \multicolumn{2}{c}{7.5 $\pm$ 0.9}   & -- \\
 $q = M_2/M_1$     & \multicolumn{2}{c}{0.96 $\pm$ 0.02 }   & --             &  \multicolumn{2}{c}{0.79 $\pm$ 0.02 }      & -- \\
 $i$ [deg]& \multicolumn{2}{c}{88.26 $\pm$ 0.02 }           & --             &  \multicolumn{2}{c}{80.7 $\pm$ 0.8 }               & -- \\
 $K$ [km/s]        &  61.22 $\pm$ 0.81 & 63.76 $\pm$ 0.85   & --             &   126.73 $\pm$ 2.0  & 160.41 $\pm$ 2.3     & -- \\
 $T$ [K]           & 10500 (fixed)     & 10412 $\pm$ 87     & --             &  8000 (fixed) & 6654 $\pm$ 112 & -- \\
 $M$ [M$_\odot$]   & 2.81 $\pm$ 0.04   & 2.69 $\pm$ 0.04    & --             &  1.76 $\pm$ 0.03   & 1.39 $\pm$ 0.03       & -- \\
 $R$ [R$_\odot$]   & 2.42 $\pm$ 0.03   & 2.34 $\pm$ 0.03    & --             &  1.68 $\pm$ 0.03   & 1.29 $\pm$ 0.04       & -- \\
 $M_{bol}$ [mag]   & 0.23 $\pm$ 0.01   & 0.33 $\pm$ 0.01    & --             &  2.21 $\pm$ 0.05   & 3.57 $\pm$ 0.07       & -- \\
 $L_B [\%]$        & 23.5 $\pm$ 0.6    & 21.6 $\pm$ 0.9     & 54.9 $\pm$ 1.3 &  68.5 $\pm$ 0.9    & 14.7 $\pm$ 0.5     & 16.7 $\pm$ 1.1  \\
 $L_V [\%]$        & 22.3 $\pm$ 0.7    & 20.5 $\pm$ 0.8     & 57.2 $\pm$ 1.7 &  63.0 $\pm$ 1.0    & 17.0 $\pm$ 0.7     & 20.0 $\pm$ 0.8  \\
 $L_R [\%]$        & 22.7 $\pm$ 0.6    & 20.9 $\pm$ 0.7     & 56.4 $\pm$ 1.0 &  60.4 $\pm$ 0.8    & 19.1 $\pm$ 0.5     & 20.5 $\pm$ 1.0  \\
 $L_I [\%]$        &     --            &    --              & --             &  55.2 $\pm$ 0.7    & 20.2 $\pm$ 0.6     & 24.6 $\pm$ 0.9  \\ \hline
\end{tabular}
\end{table*}

\begin{table*}[t]
  \caption{The parameters of visual orbit of V348 And.}
  \label{Table_AstrOrbitV348And}
  \centering
  \begin{tabular}{c c c c}
\hline \hline
Parameter       & Our solution         & \cite{Seymour2002}  & \cite{2003SerAJ.166...43O} \\
 \hline
 $p_3$ [yr]     & 86.9    $\pm$  4.3   & 330                 & 137.958 $\pm$ 1.657        \\
 $T_0$          & 2451501 $\pm$  96    & 2449900             & 2450734.4 $\pm$ 43.8       \\
 $e$            & 0.559   $\pm$  0.015 & 0.715               & 0.7238 $\pm$ 0.0096        \\
 $a$ [arcsec]   & 0.118   $\pm$  0.013 & 0.29                & 0.1527 $\pm$ 0.0008        \\
 $i$ [deg]      & 66.0    $\pm$  3.9   & 73.8                & 62.9 $\pm$ 0.3             \\
 $\Omega$ [deg] & 61.1    $\pm$  2.4   & 64.3                & 68.4 $\pm$ 1.0             \\
 $\omega$ [deg] & 139.7   $\pm$ 10.2   & 78                  & 118.14 $\pm$ 0.02          \\
 \hline
\end{tabular}
\end{table*}


After several years of observations, there was detected a photometric decrease, indicating that the
star is really an eclipsing one. And finally after many other nights of observations we found out
that its period is much longer than previously assumed, being of about 27.7~days. However, these
eclipses are very narrow (lasting about only 0.016 of the phase, which is of about 10.4~hours), but
relatively deep, of about 0.14~mag in $R$ filter.

The combined analysis of LC+RV provides us an insight into basic physical parameters of both
components. System is composed of two very similar stars. Primary and secondary components are both
of B9V spectral type. The results of our analysis are given in Table \ref{LCRVparam}, while the
plots of the RV curve as well as the light curves are given in Fig. \ref{FigRVLC_V348And}. The
third component is even more luminous than the eclipsing pair itself, and according to the spectra
its spectral type should be similar, of about B8V. As one can see, the eccentricity of the orbit is
only small, while the $\omega$ angle remains practically the same, hence the apsidal motion is only
very slow (longer than 1000~yr). 
For this eclipse-times analysis of apsidal motion we collected all available times of eclipses, our
new observed ones as well as those derived from photometry by Hipparcos \citep{HIP}, INTEGRAL/OMC
\citep{2003A&A...411L.261M}, and MASCARA \citep{2018A&A...617A..32B}. All of these are given below
in Table \ref{MINIMA}.

Thanks to its brightness and period, both the masses and radii can be derived very precisely at the
level of 2\% only. Hence, we can state that there is no other similar long-period (P$>$20d) system
with main-sequence components in our Galaxy with such well-derived parameters (see the DEBCat
catalogue by \citealt{2015ASPC..496..164S}).

Besides the photometry and spectroscopy, also the visual orbit of V348~And (i.e. WDS~J00153+4412AB)
was recomputed. Our new solution (see Table \ref{Table_AstrOrbitV348And}) differs from the already
published solutions. Both \cite{Seymour2002} and \cite{2003SerAJ.166...43O} presented much longer
orbits. But according to our modelling, there is a need for tighter orbit due to its total mass. As
one can compute from both eclipsing components, and using the Hipparcos parallax, the total mass of
the system is of about 8.1~$\mathrm{M_\odot}$. Subtracting the masses of primary and secondary, the
tertiary mass should have of about 2.6~$\mathrm{M_\odot}$. But such a small mass is unrealistic for
such a luminous body. Hence, our final conclusion is that the total mass of the system is of about
8.9~$\mathrm{M_\odot}$ (i.e. 2.8+2.7+3.4), but the system is located slightly more away, i.e.
having the parallax of 2.92~mas. Such a value is still within the error interval of the Hipparcos
value, but outside of the new GAIA parallax. We can only speculate that there can be a problem with
deriving the parallax value when the instrument is not able to resolve the double into two separate
targets and this can shift the true value of parallax.

\section{V572~Per}

The second system in our study is called V572~Per (= HIP~15193, HD~20096, $V_{max}$ = 6.7~mag). It
is also a detached binary, but it was never studied before. Its photometric variability was also
recognized by the Hipparcos \citep{HIP}, having the orbital period of about 1.21~days. Similarly to
V348~And it also contains close visual component. However, this component (being of about
1.5$^{\prime\prime}$) is not moving noticeably on the sky, hence any reliable third-body orbit
cannot be derived. Moreover, the star was also proved as a member of close open cluster Alpha
Persei \citep{2012ApJ...752...58Z}. The GAIA survey \citep{2018A&A...616A...1G} provided new
measurement of its parallax, however it gives two numbers for two close sources. The brighter one
8.07 $\pm$ 0.05~mas, while the fainter one 7.97 $\pm$ 0.06~mas (i.e. the distance 123--126~pc).

 \begin{figure*}
   \centering
   \includegraphics[width=0.99\textwidth]{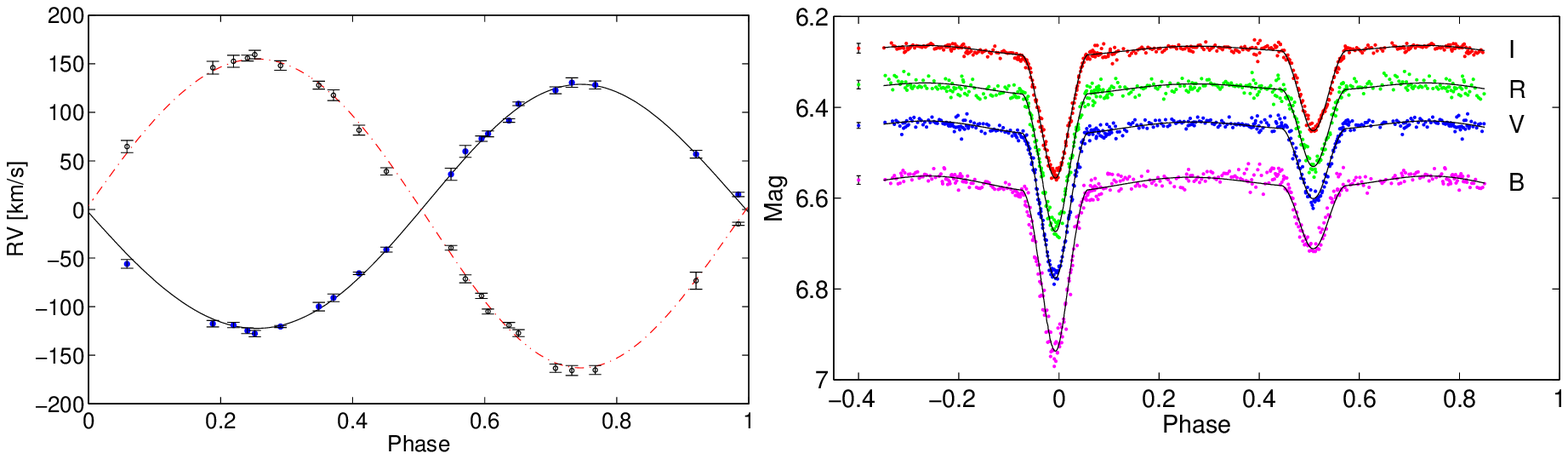}
   \caption{Result of our RV+LC fitting of V572 Per, notation same as in Fig. \ref{FigRVLC_V348And}.}
   \label{FigRVLC_V572Per}
  \end{figure*}

  \begin{figure}
   \centering
   \includegraphics[width=0.49\textwidth]{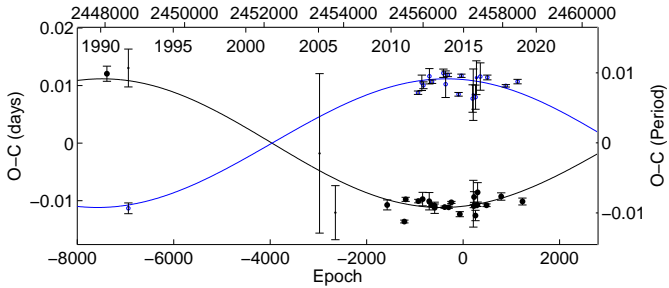}
   \caption{Variations of the minima times plotted in the $O-C$ diagram. Black solid line and the full
   dots stand for primary eclipses, while the blue line and the open circles represent the secondary ones.}
   \label{FigOC_V572Per}
  \end{figure}

Results of our LC+RV fitting are given in Table \ref{LCRVparam}, while the plots of RV and LC are
presented in Fig. \ref{FigRVLC_V572Per}.

Moreover, besides the LC+RV fitting, we also performed an analysis of the minima times of V572~Per,
which revealed that the system shows an apsidal motion of its eccentric orbit. Such a movement is
plotted in Fig. \ref{FigOC_V572Per}, where the apsidal motion during almost 30~years is clearly
visible. Its period is of about 48~years only, hence the system deserves special attention and new
photometric observations in the upcoming years to confirm this apsidal motion. Relativistic
contribution to the total apsidal motion rate is of about 4\%. An internal structure constant
resulted in log~$k_2 =-2.37$, which can be compared with the tables e.g. by
\cite{2004A&A...424..919C}. There resulted that the system is still rather young now, having the
age of about 20~Myr only.

From the combined analysis of LC+RV there also resulted that the third component in the light curve
contributes of about 20\% to its total luminosity (depending on the photometric filter). This
yielded a magnitude difference between the combined light of an eclipsing pair and this third
component of about 1.5~mag. Such a result is in perfect agreement with the values given in the
4$^{th}$ Catalog of Interferometric Measurements of Binary Stars \citep{2001AJ....122.3480H}, which
are in the range from 1.48 to 1.65~mag.

We can also compute the photometric distance to the system using our derived values. At this point
we found a problem. According to different published papers on open cluster Alpha Persei (e.g.
\citealt{2010MNRAS.402.1380J}, \citealt{1999A&A...345..471R}, or \citealt{1998ApJ...504..170P}),
its distance modulus is probably in between 6.0 and 6.5~mag (i.e. the distance 158 -- 200~pc). And
according to \cite{2006AJ....131.2967M} the star is a member of that cluster. However, as one can
check from its proper motion and distance, it is quite disputable member of such cluster. According
to our result, the distance modulus resulted in about 5.0~mag (i.e. 100~pc distant), which means
the star is much closer to the Sun than the cluster itself and is very probably not connected with
it. The value of GAIA parallax provides a distance in between these two distances.

\section{Discussion and Conclusions}

In the era of huge surveys both on photometry and spectroscopy, one can ask whether a dedicated
study on one particular binary is still worth of an effort. However, as was presented quite
recently \citep{2018ApJS..235...41K}, the catalogue of eclipsing binaries with eccentric orbits and
apsidal motion is still only sparsely populated with stars having the complete LC+RV solution
(hence with precisely derived masses) with periods longer than 20~days. Our study of V348~And can
serve as a good example. Moreover, its parameters were derived with high precision at a level of
about 2\%. The other system V572~Per can enrich a still limited group of short eccentric systems
with fast apsidal motion (U$<$50~yr), which nowadays comprises only 21 systems
\citep{2018ApJS..235...41K}. Hence any new contribution, moreover with derived masses, is welcome.
And as was mentioned in many other publications (like e.g. \citealt{2010A&A...519A..57C}) their
role for testing the stellar structure theories is still undisputable.

Moreover, besides the eccentric inner orbit, also the fact that we deal with hierarchical triple
systems is of high importance. As was presented recently e.g. by \cite{2014AJ....147...87T}, there
is an excess of tight inner binaries within the triples, probably caused by tidal evolution and
Kozai cycles.

It also should be noted that both these stars are of high brightness and also located on the
northern sky. One would expect that almost all of the interesting systems here are known and were
studied before. But, as we have pointed out, also some interesting results still can be obtained
with very small photometric instruments (all photometry for our study was obtained by using only
the small telescopes having the aperture of less than 8~cm).

 \begin{acknowledgements}
We would like to thank Ms. J.A.Nemravov\'a, Mr. J. \v{C}echura, and  Mr. R. K\v{r}\'{\i}\v{c}ek for
their contribution to the spectroscopic observations of both targets. An anonymous referee is also
to be acknowledged for his/her valuable comments improving the manuscript. This research has made
use of the Washington Double Star Catalog maintained at the U.S. Naval Observatory.  This
investigation was supported by the Czech Science Foundation grants No. P209/10/0715, GA15-02112S,
and GA17-00871S. Work is partly based on the data from the OMC Archive at CAB (INTA-CSIC),
pre-processed by ISDC. This research has made use of the SIMBAD and VIZIER databases, operated at
CDS, Strasbourg, France and of NASA's Astrophysics Data System Bibliographic Services.
 \end{acknowledgements}

\bibliographystyle{apj}
\bibliography{citace}

\section{Online-only material}

\begin{table}[t]
 \caption{Heliocentric minima of the systems used for our analysis.}
 \label{MINIMA}
 \scriptsize
 \centering
 \begin{tabular}{l l l c c l}
 \hline\hline
 Star  & HJD - 2400000 & Error  & Type & Filter & Reference\\
 name  & [d]           & [d]    &      &        &          \\
 \hline
 V348 And & 48498.65159  & 0.00627 & P & Hp & Hipparcos \\
 V348 And & 54482.66675  & 0.04519 & P & V  & OMC/INTEGRAL \\ 
 V348 And & 57142.16495  & 0.00016 & P & C  & MASCARA \\
 V348 And & 57155.82545  & 0.00834 & S & C  & MASCARA \\
 V348 And & 57252.99676  & 0.00107 & P & C  & MASCARA \\
 V348 And & 57266.64006  & 0.00222 & S & C  & MASCARA \\
 V348 And & 57308.37877  & 0.00742 & P & C  & MASCARA \\
 V348 And & 57391.48638  & 0.00540 & P & C  & MASCARA \\
 V348 And & 56269.31277  & 0.02211 & S & R  & This study \\ 
 V348 And & 56560.38780  & 0.00064 & P & V  & This study \\ 
 V348 And & 56629.45404  & 0.00092 & S & RC & This study \\ 
 V348 And & 56629.45582  & 0.00102 & S & R  & This study \\ 
 V348 And & 56906.50226  & 0.00626 & S &BVR & This study \\ 
 V348 And & 56920.54223  & 0.00290 & P & R  & This study \\ 
 V348 And & 56934.20656  & 0.00149 & S & R  & This study \\ 
 V348 And & 57017.30505  & 0.00101 & S & R  & This study \\ 
 V348 And & 57031.37299  & 0.00169 & P & C  & This study \\ 
 V348 And & 57266.63076  & 0.00192 & S & R  & This study \\ 
 V348 And & 57280.69077  & 0.00202 & P & R  & This study \\ 
 V348 And & 57294.34885  & 0.00453 & S & R  & This study \\ 
 V348 And & 57308.39410  & 0.00389 & P &BVR & This study \\ 
 V348 And & 57626.77573  & 0.00414 & S & R  & This study \\ 
 V348 And & 57696.24057  & 0.00211 & P &BVR & This study \\ 
 V348 And & 57751.65673  & 0.00088 & P & R  & This study \\ 
 V348 And & 58028.68835  & 0.00571 & P & R  & This study \\ 
 V348 And & 58042.34567  & 0.00301 & S &BVR & This study \\ 
 V348 And & 58056.38505  & 0.01229 & P & R  & This study \\ 
 V348 And & 58402.47455  & 0.00145 & S & I  & This study \\ 
 V348 And & 58416.53209  & 0.00800 & P & R  & This study \\ 
\hline
 V572 Per & 48046.84267  & 0.00120 & P & Hp & Hipparcos \\
 V572 Per & 48585.49511  & 0.00299 & P & Hp & Hipparcos \\
 V572 Per & 48586.07734  & 0.00085 & S & Hp & Hipparcos \\
 V572 Per & 53793.64709  & 0.00427 & P & V  & OMC/INTEGRAL \\  
 V572 Per & 53394.52152  & 0.01259 & P & V  & OMC/INTEGRAL \\  
 V572 Per & 57263.34037  & 0.00369 & P & V  & OMC/INTEGRAL \\  
 V572 Per & 57245.76753  & 0.00217 & S & C  & MASCARA \\
 V572 Per & 57282.75239  & 0.00084 & P & C  & MASCARA \\
 V572 Per & 57317.93136  & 0.00082 & P & C  & MASCARA \\
 V572 Per & 57318.55851  & 0.00191 & S & C  & MASCARA \\
 V572 Per & 57370.10208  & 0.00153 & P & C  & MASCARA \\
 V572 Per & 57438.66681  & 0.00216 & S & C  & MASCARA \\
 V572 Per & 55968.29287  & 0.00112 & S & C  & This study \\ 
 V572 Per & 55988.29012  & 0.00107 & P &BVRI& This study \\ 
 V572 Per & 56008.32734  & 0.00053 & S & R  & This study \\ 
 V572 Per & 56154.49526  & 0.00127 & P &BVRI& This study \\ 
 V572 Per & 56157.54991  & 0.00111 & S &BVRI& This study \\ 
 V572 Per & 56205.44812  & 0.00026 & P & R  & This study \\ 
 V572 Per & 56225.48699  & 0.00027 & S & C  & This study \\ 
 V572 Per & 56290.37051  & 0.00089 & P &BVRI& This study \\ 
 V572 Per & 56290.37017  & 0.00028 & P & I  & This study \\ 
 V572 Per & 56510.58550  & 0.00066 & S & V  & This study \\ 
 V572 Per & 56541.49831  & 0.00017 & P & V  & This study \\ 
 V572 Per & 56566.38983  & 0.00201 & S &BVRI& This study \\ 
 V572 Per & 56623.41085  & 0.00027 & S & C  & This study \\ 
 V572 Per & 56643.40532  & 0.00017 & P & C  & This study \\ 
 V572 Per & 56711.34419  & 0.00016 & P & C  & This study \\ 
 V572 Per & 56891.52002  & 0.00029 & S & C  & This study \\ 
 V572 Per & 56922.43527  & 0.00040 & P & C  & This study \\ 
 V572 Per & 56964.31399  & 0.00026 & S & I  & This study \\ 
 V572 Per & 57261.53953  & 0.00406 & S & C  & This study \\ 
 V572 Per & 57275.47175  & 0.00026 & P & V  & This study \\ 
 V572 Per & 57340.39911  & 0.00267 & S & C  & This study \\ 
 V572 Per & 57366.46036  & 0.00025 & P & C  & This study \\ 
 V572 Per & 57594.53794  & 0.00021 & P & I  & This study \\ 
 V572 Per & 57614.57759  & 0.00038 & S & V  & This study \\ 
 V572 Per & 57964.55903  & 0.00059 & P & C  & This study \\ 
 V572 Per & 58080.43684  & 0.00019 & S & R  & This study \\ 
 V572 Per & 58376.45323  & 0.00035 & S & V  & This study \\ 
 V572 Per & 58498.35689  & 0.00054 & P & V  & This study \\ 
 \hline \hline
\end{tabular}
\end{table}

\begin{table}[t]
 \caption{Radial velocities of the systems used for our analysis.}
 \label{RVs}
 \centering
 \begin{tabular}{l l r c r c}
 \hline\hline
 Star   & HJD - 2400000 & RV$_1$ & Error &   RV$_2$ & Error  \\
 name   & [d]           & [km/s] & [km/s]& [km/s]   & [km/s] \\
 \hline
 V348 And & 56041.5766 &  -69.91 & 1.65 &   50.77 &  0.39 \\
 V348 And & 56357.2585 &   37.68 & 2.01 &  -59.13 &  0.48 \\
 V348 And & 56400.6093 &  -76.20 & 1.66 &   53.01 &  7.92 \\
 V348 And & 56534.5871 &  -46.54 & 0.87 &   21.13 &  1.16 \\
 V348 And & 56563.6216 &  -61.96 & 2.66 &   37.55 &  2.48 \\
 V348 And & 56572.5450 &  -23.97 & 0.98 &    9.45 &  1.28 \\
 V348 And & 56590.2923 &  -48.02 & 1.12 &   22.40 &  4.06 \\
 V348 And & 56590.5739 &  -50.89 & 1.48 &   31.41 &  3.09 \\
 V348 And & 56592.3997 &  -72.36 & 1.36 &   52.12 &  1.45 \\
 V348 And & 56596.6362 &  -63.08 & 2.15 &   40.61 &  1.12 \\
 V348 And & 56665.1968 &   46.17 & 2.10 &  -76.87 &  1.93 \\
 V348 And & 56665.4144 &   46.11 & 2.66 &  -74.71 &  2.50 \\
 V348 And & 56862.4588 &   29.47 & 0.38 &  -55.96 &  2.47 \\
 V348 And & 56924.4092 &  -67.61 & 2.05 &   46.82 &  4.51 \\
 V348 And & 57084.2744 &   23.89 & 2.60 &  -50.82 &  3.06 \\
 V348 And & 57248.4857 &   45.88 & 7.34 &  -68.68 &  5.07 \\
 V348 And & 57260.5891 &  -69.17 & 2.55 &   48.80 &  0.80 \\
 V348 And & 57275.5356 &   46.09 & 4.10 &  -74.11 &  2.39 \\
 V348 And & 57323.5005 &    1.75 & 6.28 &  -24.11 &  6.24 \\
 V348 And & 57328.5228 &   45.15 & 1.43 &  -70.60 &  3.27 \\
 V348 And & 57868.5572 &  -76.57 & 0.80 &   54.89 &  1.93 \\
 V348 And & 57876.5621 &   -7.49 & 2.37 &    ---  &       \\
 V348 And & 57954.5020 &  -61.05 & 2.28 &   38.25 &  2.76 \\ \hline
 V572 Per & 56571.4889 &  122.65 & 3.51 & -163.36 &  4.32 \\
 V572 Per & 56572.5780 &   77.87 & 3.10 & -104.99 &  2.56 \\
 V572 Per & 56590.3078 & -118.92 & 2.72 &  152.61 &  6.22 \\
 V572 Per & 56590.4916 &  -90.88 & 3.93 &  117.60 &  5.50 \\
 V572 Per & 56592.3713 &   56.92 & 3.98 &  -73.27 &  8.75 \\
 V572 Per & 56862.5158 &   72.87 & 2.78 &  -88.93 &  2.63 \\
 V572 Per & 56862.5837 &  108.78 & 1.93 & -127.26 &  3.57 \\
 V572 Per & 56924.3582 &   59.83 & 6.17 &  -71.36 &  4.16 \\
 V572 Per & 57079.3053 & -120.45 & 0.99 &  148.18 &  4.86 \\
 V572 Per & 57084.3023 &  -65.68 & 1.30 &   81.68 &  5.05 \\
 V572 Per & 57084.3527 &  -41.35 & 2.59 &   39.24 &  3.78 \\
 V572 Per & 57126.3372 &  -56.04 & 4.50 &   64.81 &  6.40 \\
 V572 Per & 57260.6045 &  130.68 & 4.84 & -165.72 &  5.07 \\
 V572 Per & 57294.4578 &   91.56 & 1.81 & -119.09 &  2.75 \\
 V572 Per & 57323.4676 &   36.29 & 6.17 &  -39.45 &  2.40 \\
 V572 Per & 57328.5854 &  128.18 & 4.15 & -165.25 &  4.62 \\
 V572 Per & 57330.3089 & -117.55 & 3.29 &  145.97 &  6.57 \\
 V572 Per & 57332.4880 &   15.51 & 2.24 &  -14.78 &  1.54 \\
 V572 Per & 57410.4423 & -124.74 & 2.91 &  155.86 &  3.00 \\
 V572 Per & 57443.3294 &  -99.99 & 4.43 &  127.98 &  4.09 \\
 V572 Per & 57876.3165 & -127.74 & 3.29 &  159.44 &  4.39 \\
 \hline \hline
\end{tabular}
\end{table}

\end{document}